\documentclass[journal]{IEEEtran}
\usepackage{amsmath,amsfonts}
\usepackage[T1]{fontenc}
\usepackage{algorithmic}
\usepackage{array}
\usepackage{subcaption}
\usepackage{textcomp}
\usepackage{stfloats}
\usepackage{url}
\usepackage{verbatim}
\usepackage{graphicx}
\def\BibTeX{{\rm B\kern-.05em{\sc i\kern-.025em b}\kern-.08em
    T\kern-.1667em\lower.7ex\hbox{E}\kern-.125emX}}
\usepackage{balance}
\usepackage{arydshln}
\usepackage{amsmath}
\usepackage{amssymb}
\usepackage{color}
\usepackage{xspace}

\usepackage{multirow}
\usepackage{tabularx}
\usepackage{tablefootnote}
\usepackage{booktabs}
\usepackage{makecell}

\usepackage{amssymb}
\usepackage{pifont}

\usepackage{microtype}
\usepackage{graphicx}
\usepackage{booktabs}
\usepackage{hyperref}
\usepackage{enumitem}

\newcommand{\task}{novel view RIR prediction}

\newcommand{\tool}{\textsc{EigeNet}\xspace}

\begin{document}
\title{\tool: Geometry-Informed Multi-Modal Learning for Few-shot Novel View RIR Prediction}
\author{Chong Jing,~\IEEEmembership{}
        Zitong Lan,~\IEEEmembership{}
        Junan Zhang,~\IEEEmembership{}
        Zhizheng Wu~\IEEEmembership{}

\thanks{Chong Jing, Junan Zhang and Zhizheng Wu are affiliated with The Chinese University of Hong Kong, Shenzhen. e-mail: \href{mailto:225045026@link.cuhk.edu.cn}{225045026g@link.cuhk.edu.cn}}
\thanks{Zitong Lan is  affiliated with University of Pennsylvania. e-mail:\href{mailto:ztlan@upenn.edu}{ztlan@upenn.edu}}
\thanks{Zhizheng Wu is the corresponding author. e-mail: \href{mailto:wuzhizheng@cuhk.edu.cn}{wuzhizheng@cuhk.edu.cn}}
}

\maketitle

\begin{abstract}
Predicting spatially varying Room Impulse Response (RIR) from sparse observations is a critical but highly challenging inverse problem for immersive spatial audio rendering. In this work, we present \textbf{\tool}, a geometry-informed multi-modal framework for few-shot novel view RIR prediction. At its core is a Cross-view Alternate-attention Transformer that iteratively refines local intra-view acoustic structures and global cross-view spatial relationships. We empirically demonstrate that this architecture is capable of making full use of the multi-view multi-modal context while performing spatial-temporal reasoning for RIR prediction.
Inspired by acoustic ray tracing, we design a geometry-informed  modulation block to formulate the connection between geometric features and RIR power spectrum.
In the mean time, an auxiliary loss is introduced to transform the single-target waveform prediction into a multi-task learning framework. 
Through ablation studies, we demonstrate that this design yields consistent performance gains regardless of the underlying backbone, thereby confirming its foundational utility and architecture-agnostic generalizability for RIR prediction task.
Evaluated on both simulated and real-world benchmarks, \tool achieves both state-of-the-art performance in few-shot novel view RIR prediction and sim-to-real generalization. Codes and checkpoints are available on \url{https://github.com/FEAfeatherTHER/EigeNet}.

\end{abstract}


\section{Introductio}
Immersive interactions within spatial computing, such as Augmented/Virtual Reality (AR/VR)~\cite{nerf, gaussianssplatting} and generative world models~\cite{lingbot, genie}, fundamentally rely on the seamless integration of both visual and auditory modalities. While novel view prediction for visual environments has reached a highly mature state, the accurate construction of corresponding acoustic fields remains a critical yet underexplored domain. The auditory identity of an environment is essentially encoded within its Room Impulse Response (RIR), which captures complex acoustic phenomena ranging from echoes and spatial reverberation to the overall spatial acoustics of the environment~\cite{fdtd1,fdtd2,kwave}.

Physically, the acoustic field of the room functions as a highly complex dynamical system, where the RIR denotes the precise spatio-temporal response at a receiver location to a Dirac impulse applied at a source location. As acoustic waves radiate, they undergo complex reflection, diffraction, scattering, and transmission against the physical boundaries of the room~\cite{rt1,rt2,pbr}. These interactions produce spatially varying acoustic field throughout the room so people can have distinct auditory experience. Due to the complexity of these interactions, acquiring high-fidelity acoustic fields traditionally demands exhaustive and resource-intensive physical measurements~\cite{recording, raf}.

To circumvent the data dependency of demanding physical measurements, recent efforts have increasingly turned to deep learning for accurate acoustic field estimation. Researchers have explored various paradigms to reconstruct acoustic patterns of a single room~\cite{naf, inras, HAA, dar,avr, versa, masuyama2025physics, ick2025direction, avnerf, avrir, NVAS, lan2025building}. However, early data-driven approaches typically focus on environment-specific modeling, inherently lacking transferability to novel scenes. Consequently, to enable scalable real-world applications, there is an urgent need to advance toward cross-room few-shot \task{}. This paradigm aims to generalize spatially varying acoustic effects to unseen environments using extremely limited reference data and room geometric information.
So far, a number of contemporary works follow this paradigm and achieve promising performance~\cite{fewshotrir, xRIR, FLAC}.
Among these, Few-shot RIR~\cite{fewshotrir} pioneer on this task by integrating multi-modal inputs within a transformer-based autoencoder, while xRIR~\cite{xRIR} formulates target RIR prediction as a weighted combination of reference-view samples. FLAC~\cite{FLAC} is the first to employ a generative framework for this task.

Despite these efforts on generalizing RIR prediction to unseen environments, existing methods still struggle to achieve high-fidelity acoustic field estimation due to inherent constraints in their architectural and optimization designs.
Specifically, current methods are fundamentally constrained by two unresolved bottlenecks. \emph{The first challenge is the difficulty in simultaneously capturing the temporal structure within a single RIR and the spatial relationships among reference observations.}
Existing approaches mostly rely on Self-attention or Cross-attention to implicitly learn these temporal-spatial relationships.
However, standard Self-attention and Cross-attention mechanisms are not explicitly designed for multi-view multi-modal contexts~\cite{vggt, fastvggt, wang2025flashvggtefficientscalablevisual}. 

\emph{Another challenge is that most existing methods treat the task as a black-box mapping.} Failing to explicitly model the inherent multi-modal correlations between room geometry and acoustic patterns, existing works often lack physical interpretability and struggle with complex geometric configurations.

To overcome these non-trivial challenges, we introduce a novel framework \tool that tackles these bottlenecks.
Firstly, we introduce a \emph{Cross-view Alternate-attention transformer (CVAT)}, alternating between intra-view (local) and cross-view (global) attention. 
The efficacy of Alternate-attention has already been well-established in the domain of novel-view image reconstruction~\cite{vggt, fastvggt, wang2025flashvggtefficientscalablevisual}. However its application to audio-centric multi-modal learning is still unknown, so we bridge this gap by presenting the first framework that leverages Alternate-attention for novel-view RIR prediction. 
Beyond demonstrating state-of-the-art performance, we also conduct an in-depth analysis to explore the inner mechanism that drive this success.

To tackle with the second challenge, we draw inspiration from the principles of physics-based acoustic ray tracing~\cite{pbr, roomacoustics} that the multi-octave power spectrum of RIR can be estimated from the room geometry and surface material properties.
Leveraging this deterministic relationship as a powerful physical prior, we introduce a geometry-informed modulation block to explicitly modulate acoustic representations using geometric features.

Additionally, to maximize the efficacy of this constraint, we incorporate an auxiliary loss on power spectrum, thereby transitioning the conventional single-target prediction into a multi-task learning paradigm. This design significantly enhances the cross-room generalizability of the network while ensuring the physical plausibility of the synthesized RIRs.

We conducted extensive experiments to validate the efficacy of these components. We first demonstrate the fundamental distinctions in the inference paradigms between Alternate-attention and other attention mechanisms when attending multi-view multi-modal context. Subsequently, we elucidate that the introduction of the modulation block and corresponding auxiliary loss exhibits strong generalizability, as it yields consistent performance gains across various network architectures, regardless of the underlying attention mechanism employed.

In summary, our primary contributions are as follows:
\begin{itemize}
    \item We propose a Cross-view Alternate-attention transformer into few-shot \task{} and explain its efficiency over Self-attention and Cross-attention when dealing with multi-view multi-modal context.
    \item We design a geometry-informed modulation block to strengthen the correlation between geometric and acoustic features.
    \item We introduce an auxiliary loss, shifting the task from single-target RIR prediction to multi-modal multi-task learning paradigm and empirically validate its effectiveness.
\end{itemize}
Our proposed model achieves state-of-the-art (SOTA) performance on AcousticRooms and Hearing-Anything-Anywhere datasets, while exhibiting superior robustness even with sparse references.

\section{Related Work}
\subsection{Few-shot Novel View RIR Prediction with Multi-Modal Methods}
Few-shot Novel View RIR prediction task tries to learn the room acoustic patterns and predict the target RIR in the same room. In multi-modal learning field, there are three related excellent works. Few-shot RIR~\cite{fewshotrir} utilizes RGB, depth, RIR and other metadata to predict the log magnitude spectrogram with a cross-modality Unet architecture. 
xRIR~\cite{xRIR} introduces a geometric module and views the target RIR as the weighted sum of the reference RIRs.
FLAC~\cite{FLAC} is built upon the Latent Diffusion Model (LDM)~\cite{chen2023pixartalphafasttrainingdiffusion, liu2023audioldmtexttoaudiogenerationlatent, esser2024scalingrectifiedflowtransformers} and incorporates the AGREE joint acoustic-geometric embedding like CLIP~\cite{radford2021learningtransferablevisualmodels}, which is specifically designed for robust feature extraction across both acoustic and geometric modalities.

\subsection{Differentiable rendering of Acoustic Field}
Differentiable rendering methods are effective and popular because they combine physical acoustic knowledge with neural networks. Several representative works have emerged along this line.
AVR~\cite{avr} adopts the idea from NeRF and derives the rendering equation in the frequency domain, but it cannot handle cross-room scenarios. Diff-RIR~\cite{HAA} adopts the image-source method and builds a complete and detailed pipeline; however, it is resource-consuming due to the need to filter out invisible image sources, and it does not model frequency-dependent responses. DAR~\cite{dar} is the first to combine ray-tracing results with visual features and achieves strong performance. Although differentiable rendering methods have achieved considerable success in RIR prediction, they still have to rely on physical approximations that may not hold in all situations, which is a theoretical limitation of this paradigm.

\subsection{Alternate-attention}
Visual Geometry Grounded Transformer (VGGT) first introduces Alternate-attention as a 3D feature extractor and evaluates its effectiveness on the downstream tasks like novel view image prediction\cite{vggt, ff3d_review,fastvggt}.
An Alternate-attention layer is composed of a frame wise and a global wise Self-attention layer, so the attention will be focused alternately on the frame and global wise context. Alternate-attention is specially designed for multi-view context and VGGT uses experiments to prove its superiority over Cross-attention and Self-attention (Global Self-attention only) structures\cite{vggt}.
While existing VGGT-style Alternating Attention (Alternate-attention) architectures are exclusively tailored for pure 3D visual tasks, Room Impulse Responses (RIRs) fundamentally consist of 1D temporal signals coupled with cross-view spatial configurations, exhibiting distinct token structures and attention symmetries. To the best of our knowledge, our work represents the first attempt to adapt the Alternate-attention mechanism to such multi-modal multi-view context.

\section{Method}
\subsection{Problem Definition}

Following xRIR's setup~\cite{xRIR}, we are given the source and receiver data: position coordinates of sources ${tx}_i\in \mathbb{R}^{3}$ and receivers ${rx}_i\in \mathbb{R}^{3}$ in the camera coordinate system (setting the receiver as the origin like xRIR). 
At these reference pairs, we have reference RIRs $h_i\in \mathbb{R}^{1\times L}$ ($T$ denotes the sample length in time).
We also have a panorama depth map at receiver location from its viewpoint, denoted as ${D}$.
Given these information, our goal is to estimate the target RIR $h_0$ at novel source position ${tx}_0$. We formulate this as follows:
\begin{equation}
h_0 = \mathcal{F}\big(\{h_i, tx_i, rx_i\}_{i=1}^N,\, tx_0,\, D \big),
\end{equation}
where $\mathcal{F}$ denotes our model \tool.

\begin{figure*}
    \centering
    \includegraphics[width=0.98\linewidth]{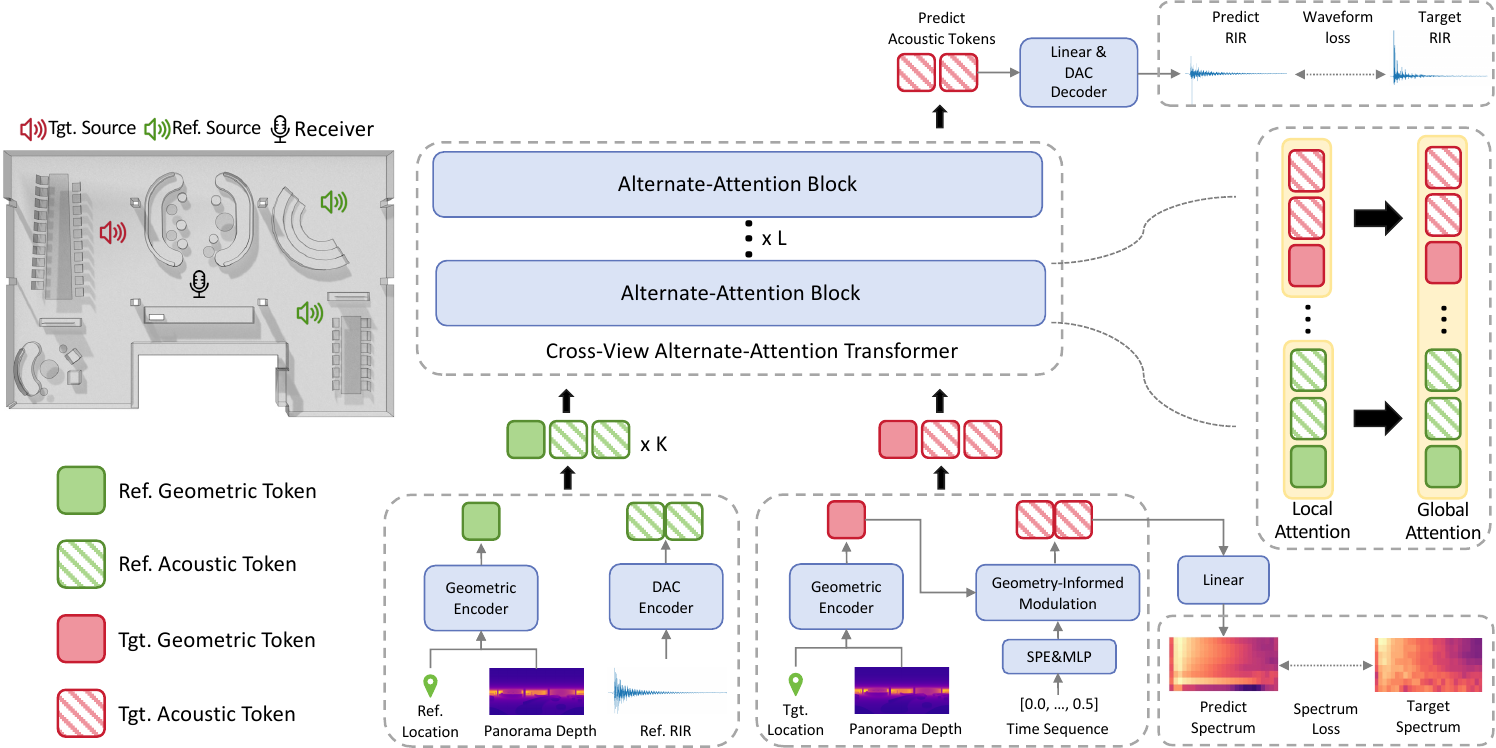}
    \caption{Overview of \tool. Geometric and acoustic tokens of all reference and target views are firstly encoded by Encoders and a modulation block, then jointly processed by the Cross-view Alternate-attention Transformer (CVAT) that interleaves intra-view and cross-view attention to predict the target RIR.}
    \label{fig:network}
\end{figure*}

\subsection{Model Overview}
As shown in Fig~\ref{fig:network}, we employ a geometric encoder to encode the panoramic depth map and position coordinates into geometric tokens. Then we have two different branches to get the acoustic tokens of the reference and target view respectively.
Taking geometric token as the condition, the geometry-informed modulation block modulate the target RIR acoustic tokens and regress those tokens with the multi-octave power spectrum.
After that, we concatenate geometric token before its corresponding acoustic tokens for both the reference and target views and pass the tokens into our Cross-view Alternate-attention Transformer (CVAT) backbone. 
Finally, the target acoustic tokens are decoded into waveform to predict the target RIR.
We explain next the details of our model architecture.

\subsection{Encoders}

\subsubsection{Geometric} 
We generally follow similar design of xRIR~\cite{xRIR}. 
Within the camera coordinate system centered at the receiver, we first project the depth map back to 3D pixel coordinates using an isometric mapping. To characterize the sound propagation paths, we compute two tensors by subtracting the source and receiver coordinates from the 3D pixel coordinates, respectively.
We concatenate these two tensors into a single 6-channel tensor then patchify into a sequence of patches and convert to tokens with a standard Vision Transformer (ViT)~\cite{vit} model.
For the coordinates encoding, we apply positional encoding followed by a multi-layer perceptron (MLP) layer. 
Finally, the resulting components from the coordinates and the depth map are concatenated and projected to match the dimensions of the acoustic tokens.
The resulting geometry tokens is $G_i\in R^{1\times f}$ ($i \in [0,...,N]$) for all the reference views geometry and target view geometry. 
We keep the feature dimension of the geometric token the same with acoustic tokens so that they can be concatenated and jointly processed by the CVAT backbone.

\subsubsection{Acoustic}
The Encoders for acoustic tokens are designed with two distinct branches to independently process the reference views and the target view.

For reference RIR, we adopt Descript-Audio-Codec~\cite{dac}, which is a widely adopted tokenizer in the audio domain, capable of reconstructing audio with high fidelity acoustic details.
Considering the domain shift from general audio to RIR signal, we empirically test the DAC reconstruction performance by evaluating RIR reconstruction metrics on the AcousticRooms dataset~\cite{xRIR}. 
We will provide the preliminary reconstruction quality in Sec.~\ref{sec:exp}.
Given the exceptionally high reconstruction quality, we adopt the pretrained 16~kHz version of DAC without any finetuning. 
Specifically, we extract the continuous latent representations prior to the vector quantizer~\cite{vqvae} as acoustic tokens $A_i\in \mathbb{R}^{n\times f}$, where $n$ and $f$ are the number of tokens and the dimension of the acoustic features, $i \in [1,...,N]$ denotes N reference views.

For the target view RIR, we utilize the sinusoidal positional encoding (SPE)~\cite{attention} of its time sequence as a proxy and project them into a high-dimensional space $T_0\in \mathbb{R}^{n\times f}$ with a MLP, where the subscript $0$ denotes the index for the target view. $T_0$ will then be modulated by the geometry-informed modulation block to get the acoustic tokens, which we will explain next.

\subsection{Geometry-informed Modulation Block.}
Geometry-informed Modulation Block is designed to explicitly strengthen the correlation between the room geometry and power spectrum of RIR.

In ray-tracing algorithms, multi-octave power spectrum can be analytically derived from room geometry and surface acoustic properties. We identify this as an inherent multimodal correlation that has been largely under-explored in prior data-driven methods.
Leveraging this insight, we modulate $\mathbf{T}_0$ by its corresponding geometric token $G_0$ via a one-layer Diffusion Transformer (DiT) block with adaptive layer normalization\cite{dit}, yielding the acoustic tokens $\mathbf{A}_0 \in \mathbb{R}^{n\times f}$ for target view.
Subsequently, $\mathbf{A}_0$ is divided into two streams: one is passed into CVAT backbone, while the other is projected to $\hat{\mathbf{S_0}} \in \mathbb{R}^{n\times b}$ through a linear layer to regress the power spectrum. 
Instead of regressing the original Short-time Fourier Transform (STFT) power spectrum of the targer RIR $\mathbf{Z_0} \in \mathbb{R}^{n\times B}$, where $B$ denotes the frequency bins of the spectrum, we choose to regress the multi-octave power spectrum $\mathbf{S_0} \in \mathbb{R}^{n\times b} $. In our configuration, the number of octave bands is set to $b = 7$, spanning center frequencies from 63 Hz to 4000 Hz according to the ISO standard (i.e., 63 Hz, 125 Hz, 250 Hz, 500 Hz, 1 kHz, 2 kHz, and 4 kHz).

To prevent temporal collapse, we adopt a combination of multi-resolution STFT (MRSTFT) loss and energy decay curve (EDC) loss:
\begin{equation}
\mathcal{L}_{\text{spectrum}} = \mathcal{L}_{\text{MRSTFT}}(\hat{\mathbf{S_0}}, \mathbf{S_0}) + \mathcal{L}_{\text{EDC}}(\hat{\mathbf{S_0}}, \mathbf{S_0}),
\end{equation}

\subsection{Cross-view Alternate-attention Transformer.}

\subsubsection{Token Organization}
We assemble the geometric and acoustic tokens for every viewpoint by concatenating the geometric tokens $G$ before its corresponding acoustic tokens $A$ as prefix along the sequence dimension:
\begin{equation}
\mathcal{V}_i = \text{Concat}\{\mathbf{G}_i,  \mathbf{A}_i\},
\end{equation}
where $i = 0,1,\dots,N$.
This results in tokens for each view with shape of $V_i\in R^{(1+n)\times f}$.
Following the previous definition, we let the subscripts $i \in \{1, \dots, N\}$ represent the set of reference view tokens, with $i=0$ designating the target view tokens. 
We then stack all view tokens into a unified sequence:
\begin{equation}
\mathbf{H}^{(0)} = \text{Concat}\{{\mathcal{V}_0, \mathcal{V}_1, \dots, \mathcal{V}_N}\},
\end{equation}
where $\mathbf{H}^{(0)} \in \mathbb{R}^{(N+1)(1+n)\times f}$.

\subsubsection{Alternate-attention for Acoustic Modeling}
Each view token $\mathcal{V}_i$ contains both geometric and acoustic information, where the acoustic tokens exhibit strong temporal structure within each view, while the target RIR requires aggregating information across different views under varying geometric configurations. To explicitly model these two complementary dependencies, we design a transformer that alternates between \emph{intra-view} (local) and \emph{cross-view} (global) attention.

\subsubsection{Local Attention (Intra-View)}
Local attention is applied independently within each view token $\mathcal{V}_i$, restricting attention to tokens belonging to the same view. This preserves the temporal structure of acoustic tokens while allowing interaction with the corresponding geometric tokens.
Formally, for the $\ell$-th layer, we partition $\mathbf{H}^{(\ell-1)}$ into ${\mathcal{V}_i^{(\ell-1)}}$ and apply multi-head Self-attention within each view:
\begin{equation}
\tilde{\mathcal{V}}_i^{(\ell)} = 
\mathrm{MSA}_{\text{local}}\left(\mathcal{V}_i^{(\ell-1)}\right) + \mathcal{V}_i^{(\ell-1)},
\quad i\in {0, \dots, N}.
\end{equation}
The updated tokens are then concatenated back:
\begin{equation}
\tilde{\mathbf{H}}^{(\ell)} = \text{Concat}\{\tilde{\mathcal{V}}_0^{(\ell)}, \dots, \tilde{\mathcal{V}}_N^{(\ell)}\}.
\end{equation}

\subsubsection{Global Attention (Cross-view)}
Global attention is applied over the entire token sequence, enabling information exchange across all views. This allows the model to aggregate acoustic cues from reference views to infer the target view.

\begin{equation}
\mathbf{H}^{(\ell)}=
\mathrm{MSA}_{\text{global}}\left(\tilde{\mathbf{H}}^{(\ell)}\right) + \tilde{\mathbf{H}}^{(\ell)}.
\end{equation}

\subsubsection{Alternating Structure}
We stack multiple Alternate-attention blocks, where each block consists of a local attention layer followed by a global attention layer. This alternating design enables the model to iteratively refine intra-view representations (capturing local acoustic structure), and integrate cross-view information (capturing spatial relationships).

\subsection{Prediction Head}
We extract and concatenate the local and global features from the final Alternate-attention block on the feature dimension, and project them back to the hidden space of the DAC latent representation:
\begin{equation}
\hat{\mathbf{Z}}_0 = \phi_{\text{proj}}(\mathbf{A}_0),
\end{equation}
where $\phi_{\text{proj}}$ denotes a linear projection.
Instead of regression in the latent domain, we feed the predicted latent $\hat{\mathbf{Z}}_0$ into the frozen DAC decoder to obtain the reconstructed waveform.
\begin{equation}
\hat{h}_0 = \mathrm{Decoder}(\hat{\mathbf{Z}}_0).
\end{equation}
This head directly supervises the RIR signal in the waveform domain, enabling the model to capture fine-grained temporal and spectral characteristics.

\section{Experiments}

\subsection{Dataset.} 
\subsubsection{AcousticRooms(AR)~\cite{xRIR}}
It is a large high-quality simulated dataset with 300K mono channel RIR in 22050Hz sampling rate computed on Treble\footnote{https://www.treble.tech/} platform, featuring 260 rooms across 10 categories and 332 materials across 11 categories. In our experimental setup, we downsample the RIR data to a sampling rate of 16 kHz to match with DAC.

\subsubsection{Hearing-Anything-Anywhere (HAA)~\cite{HAA}}
The HAA dataset consists of real-recorded RIRs, each featuring a fixed source and multiple receiver positions. This configuration is opposite to the AR dataset but doesn't affect the experiment due to the reciprocity of the wave equation. The original RIRs are recorded at 48 kHz and are downsampled to 16 kHz for our experiments.

\begin{table}[h]
    \centering
    \caption{Reconstruction quality of the pretrained 16~kHz Descript-Audio-Codec (DAC) on the AcousticRooms dataset. Lower is better ($\downarrow$).}
    \label{tab:dac_reconstruction}
    \resizebox{0.65\columnwidth}{!}{%
    \begin{tabular}{lccc}
    \toprule
    & \textbf{EDT (s)} $\downarrow$ & \textbf{C50 (dB)} $\downarrow$ & \textbf{T60} $\downarrow$ \\
    \midrule
    DAC & 0.004 & 0.606 & 4.963 \\
    \bottomrule
    \end{tabular}
    }
    \end{table}

\subsection{Metrics.} Following prior works~\cite{xRIR,fewshotrir}, we use three common metrics:
\begin{itemize}
    \item \textbf{EDT.} Early Decay Time Error measures the time taken for the initial 5dB decay in the energy curve.
    \item \textbf{C50.} Clarity measures the ratio of the early energy(50ms) compared to late energy.
    \item \textbf{T60.} T60 measures the time required for the cumulative energy attenuation of 60~dB after an initial 5~dB drop. Following xRIR~\cite{xRIR}, we compare the T60 value of the predicted RIR and the ground-truth RIR. Instead of computing it directly, we estimate T20 and multiply it by a factor of 3 to approximate T60 for numerical stability.
\end{itemize}
To assess the suitability of the Descript-Audio-Codec (DAC) for the downstream task of RIR prediction, we conducted an extensive evaluation across the AR dataset. As shown in Tab. \ref{tab:dac_reconstruction}, the metrics demonstrate the exceptional suitability of DAC as tokenizer for RIR prediction.

\subsection{Baselines.} Given the same input configuration, we compare our model with following baselines. Note that FLAC~\cite{FLAC} is a concurrent work to ours, and therefore is not included in our baseline comparisons.
\begin{itemize}
    \item \textbf{Random Across Rooms}: randomly sampled from the entire dataset.
    \item \textbf{Random Same Room}: randomly selected from the same room.
    \item \textbf{Linear Interp.}: linearly interpolates between K reference RIRs based on their distances to the target.
    \item \textbf{Nearest Neighbor (KNN)}: chooses the RIR among the K reference samples with the closest distance to the target.
    \item \textbf{xRIR}~\cite{xRIR}: we adopt xRIR as our primary learning-based baseline, as it has been shown to outperform earlier multi-modal approaches (e.g., Few-shot RIR~\cite{fewshotrir}) by a large margin on both AR and HAA.
    For a fair comparison, we retrain xRIR on AR dataset at 16~kHz for 10 epochs using its released configuration and scripts.
    \item \textbf{Diff-RIR}~\cite{HAA}: it introduces the HAA dataset and is a physics based differentiable rendering method.
    We only compare with it on HAA.
\end{itemize}

\subsection{Implementations.} In our model, we predict 16~kHz RIR with a length of 8000 samples (results in 0.5s in duration). The DAC tokenizer, operating at a sampling rate of 16 kHz, features a frame rate of 50 Hz. Consequently, each RIR yields 25 acoustic tokens with 1024 dimension, which serve as the latent representation for subsequent processing.
Every depth map is projected into a 6 channel tensor with $(256, 512)$ in size. 
We then divide this tensor into  $16 \times 32$ patches and input it into a 8-head, 4-layer Vision Transformer with a feature dimension of 512.
The coordinates are encoded by a positional embedding module with a hidden dimension of 128 and be projected into 512 dimension by a MLP.
Cross-view Alternate-attention Transformer has 6 Alternate-attention blocks with feature dimension of 1024 and 16 heads.
The overall architecture comprises 132.54 M parameters.
Following the former works~\cite{xRIR, inras, avr}, we also use MRSTFT loss and EDC loss for waveform supervision.
Together with the power spectrum loss, The total loss are:
\begin{equation}
\mathcal{L}_{total} = \mathcal{L}_{MRSTFT} + \mathcal{\lambda}_{EDC}\mathcal{L}_{EDC} + \mathcal{\lambda}_{spectrum}\mathcal{L}_{spectrum}
\end{equation}
where we set $\mathcal{\lambda}_{EDC} = 1$ and $\mathcal{\lambda}_{spectrum} = 0.01$.
The waveform EDC loss and power spectrum loss weights are linearly warmed up from 0 to their final values over the first 2000 training steps. We train our model for 10 epochs on 8 H100 Nvidia GPUs with a batchsize of 48 with Adam Optimizer\cite{kingma2017adammethodstochasticoptimization}.

\begin{table}[t]
\centering
\caption{Quantitative comparison on the AcousticRooms dataset under varying numbers of reference views $K\!\in\!\{1,4,8\}$. Metrics are EDT (s), C50 (dB) and T60(\%) errors; lower is better ($\downarrow$). The best result in each $K$ block is shown in \textbf{bold}.}
\label{tab:acousticrooms}
\resizebox{0.95\columnwidth}{!}{%
\begin{tabular}{lcccc}
\toprule
\textbf{Method} & \textbf{$K$} & \textbf{EDT} $\downarrow$ & \textbf{C50} $\downarrow$ & \textbf{T60} $\downarrow$ \\
\midrule
Random Across Rooms & -- & 0.442 & 7.784 & 36.420 \\
Random Same Rooms   & -- & 0.207 & 5.984 & 15.374 \\
\midrule
Nearest Neighbor (KNN) & 1 & 0.303 & 4.013 & 27.583 \\
xRIR~\cite{xRIR}       & 1 & 0.076 & 2.124 & 12.617 \\
\textbf{Ours}          & 1 & \textbf{0.052} & \textbf{1.488} & \textbf{10.213} \\
\midrule
Linear Interp.         & 4 & 0.281 & 4.552 & 17.583 \\
Nearest Neighbor (KNN) & 4 & 0.185 & 3.687 & 24.194 \\
xRIR~\cite{xRIR}       & 4 & 0.054 & 1.540 & 10.052 \\
\textbf{Ours}          & 4 & \textbf{0.047} & \textbf{1.398} & \textbf{8.061} \\
\midrule
Linear Interp.         & 8 & 0.245 & 4.078 & 25.713 \\
Nearest Neighbor (KNN) & 8 & 0.157 & 3.455 & 20.562 \\
xRIR~\cite{xRIR}       & 8 & 0.050 & 1.442 & 9.393 \\
\textbf{Ours}          & 8 & \textbf{0.041} & \textbf{1.242} & \textbf{7.605} \\
\bottomrule
\end{tabular}%
}
\end{table}

\subsection{Experimental Results and Analysis}
\label{sec:exp}
To mitigate the impact of stochastic variations, we perform three independent trials, each initialized with a different random seed, and report the average across trials for all results below.
\subsubsection{Performance on AcousticRooms Dataset} 
\begin{figure*}[t]
    \centering
    \includegraphics[width=\linewidth]{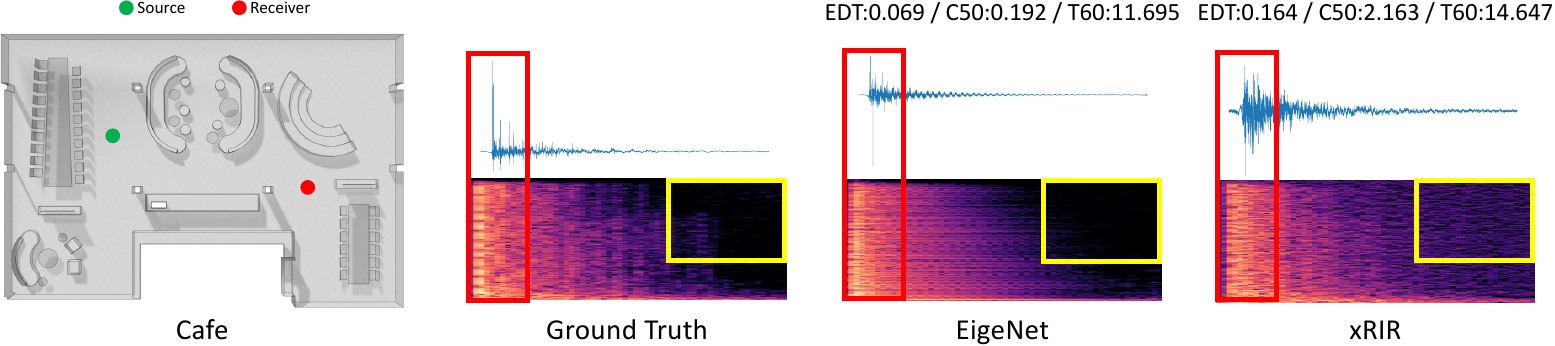}
    \caption{Qualitative results on a complex \emph{Cafe} scene from AcousticRooms: Red and yellow bounding boxes highlight the regions selected for direct comparison: the early-reflection region in the waveform and the high-frequency portion of the late tail in the STFT.}
    \label{fig:casestudy}
\end{figure*}

As shown in Tab~\ref{tab:acousticrooms}, our method consistently outperforms both learning-based (xRIR) and classical interpolation baselines across all metrics and $K\!\in\!\{1,4,8\}$. Even in the extremely sparse setting ($K{=}1$), our model achieves the lowest error, indicating strong generalization with very limited observations. Classical baselines such as linear interpolation degrade significantly under sparse settings and remain suboptimal even with more references, highlighting their inability to capture complex acoustic propagation. A detailed analysis of how performance scales with $K$ is deferred to Sec.~\ref{sec:k_scaling}.

Fig.~\ref{fig:casestudy} further provides a qualitative illustration on one challenging \emph{Cafe} scene. In the waveform red bounding box, the peak locations and relative amplitudes of the early reflections produced by \tool closely track the ground truth, whereas xRIR smears the onset and misplaces several early peaks. In the STFT yellow bounding box (late tail, high-frequency region), our prediction preserves a clean decay structure with markedly fewer high-frequency artifacts. These observations show that \tool produces outputs that are visually and structurally consistent with the ground truth across the whole time--frequency domain, which is in line with the quantitative gains observed in Tab~\ref{tab:acousticrooms}.

\begin{table*}[h]
\centering
\caption{Sim-to-real performance on the HAA dataset across all four acoustic environments. All scenes are partitioned into train and test splits with an 8:2 ratio. Results are reported under varying reference counts $K\!\in\!\{1,4,8\}$, alongside the physics-based Diff-RIR baseline at $K{=}12$. Metrics include EDT(s), C50(dB), and T60(\%) errors (lower is better, $\downarrow$). The best value in each column within a $K$ block is highlighted in \textbf{bold}.}
\label{tab:haa_metrics}
\resizebox{\textwidth}{!}{%
\begin{tabular}{lccccccccccccc}
\toprule
\multirow{3}{*}{\textbf{Method}} & \multirow{3}{*}{\textbf{$K$}} & \multicolumn{12}{c}{\textbf{All Rooms}} \\
\cmidrule(lr){3-14}
 & & \multicolumn{3}{c}{\textbf{Classroom}} & \multicolumn{3}{c}{\textbf{Hallway}} & \multicolumn{3}{c}{\textbf{Complex}} & \multicolumn{3}{c}{\textbf{Dampened}} \\
\cmidrule(lr){3-5} \cmidrule(lr){6-8} \cmidrule(lr){9-11} \cmidrule(lr){12-14}
 & & EDT $\downarrow$ & C50 $\downarrow$ & T60 $\downarrow$ & EDT $\downarrow$ & C50 $\downarrow$ & T60 $\downarrow$ & EDT $\downarrow$ & C50 $\downarrow$ & T60 $\downarrow$ & EDT $\downarrow$ & C50 $\downarrow$ & T60 $\downarrow$ \\
\midrule
Random Across Room & -- & 0.355 & 6.842 & 32.150 & 0.412 & 7.205 & 35.660 & 0.450 & 8.120 & 40.230 & 0.521 & 9.243 & 255.407 \\
Random Same Room   & -- & 0.185 & 5.237 & 14.800 & 0.205 & 5.850 & 16.207 & 0.240 & 6.554 & 18.904 & 0.285 & 7.153 & 122.823 \\
\midrule
KNN                & 1  & 0.137 & 3.522 & 25.100 & 0.185 & 3.782 & 26.814 & 0.185 & 4.152 & 28.543 & 0.125 & 4.851 & 52.532 \\
xRIR               & 1  & 0.045 & 0.994 & 12.067 & 0.043 & 0.942 & 2.829  & 0.032 & 0.801 & 12.581 & 0.047 & 4.771 & 193.582 \\
\textbf{Ours}      & 1  & \textbf{0.039} & \textbf{0.782} & \textbf{3.066} & \textbf{0.042} & \textbf{0.576} & \textbf{2.516} & \textbf{0.024} & \textbf{0.547} & \textbf{2.620} & \textbf{0.026} & \textbf{1.935} & \textbf{49.090} \\
\midrule
Linear Interp.     & 4  & 0.233 & 3.792 & 23.026 & 0.311 & 4.149 & 25.062 & 0.388 & 5.021 & 27.896 & 0.310 & 5.628 & 52.596 \\
KNN                & 4  & 0.156 & 3.413 & 20.596 & 0.162 & 3.749 & 22.087 & 0.174 & 3.980 & 24.506 & 0.184 & 4.872 & 48.169 \\
xRIR               & 4  & \textbf{0.034} & \textbf{0.798} & \textbf{3.031}  & 0.042 & 0.964 & 3.113  & 0.034 & 1.027 & 12.247 & 0.044 & 4.699 & 191.347 \\
\textbf{Ours}      & 4  & 0.038 & 0.816 & 3.227 & \textbf{0.038} & \textbf{0.557} & \textbf{2.537} & \textbf{0.022} & \textbf{0.527} & \textbf{2.717} & \textbf{0.026} & \textbf{1.925} & \textbf{46.072} \\
\midrule
Linear Interp.     & 8  & 0.214 & 3.856 & 22.437 & 0.235 & 4.124 & 24.511 & 0.264 & 4.883 & 27.941 & 0.295 & 5.506 & 46.623 \\
KNN                & 8  & 0.145 & 3.212 & 19.798 & 0.155 & 3.352 & 21.206 & 0.171 & 3.652 & 23.422 & 0.180 & 4.233 & \textbf{34.413} \\
xRIR               & 8  & 0.047 & 1.058 & 12.195 & 0.038 & 0.840 & 3.088  & 0.033 & 0.947 & 12.480 & 0.043 & 4.656 & 191.392 \\
\textbf{Ours}      & 8  & \textbf{0.037} & \textbf{0.842} & \textbf{3.463} & \textbf{0.036} & \textbf{0.519} & \textbf{2.440} & \textbf{0.023} & \textbf{0.562} & \textbf{2.764} & \textbf{0.025} & \textbf{2.002} & 49.261 \\
\midrule
Diff-RIR           & 12 & 0.065 & 2.155 & 15.396 & 0.062 & 1.951 & 14.234 & 0.058 & 2.451 & 16.826 & 0.085 & 4.253 & 85.314 \\
\bottomrule
\end{tabular}
}
\end{table*}

\begin{figure}[t]
    \centering
    \includegraphics[width=\linewidth]{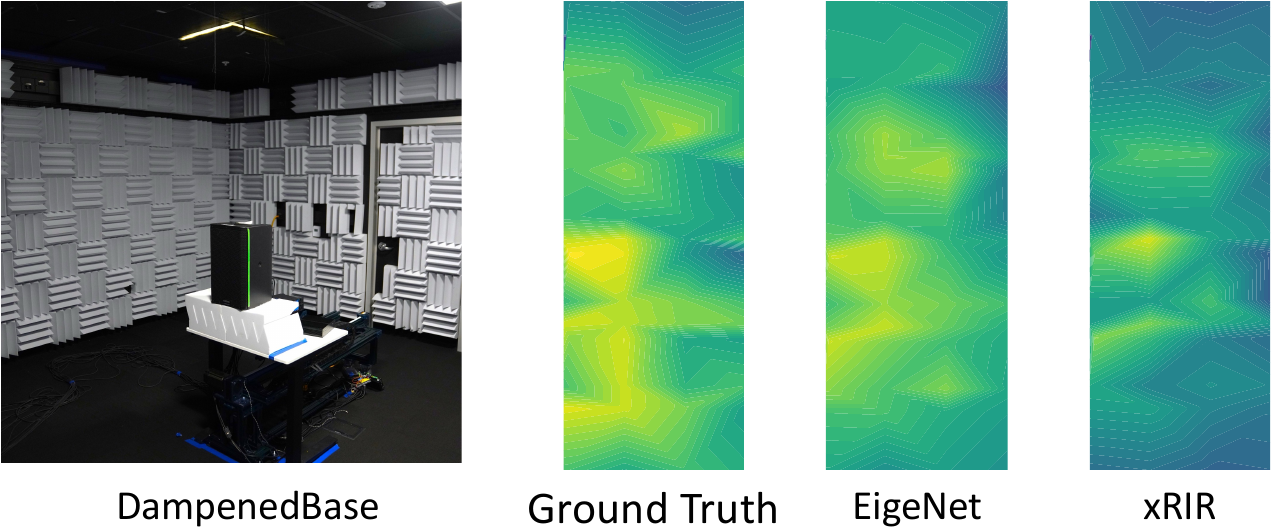}
    \caption{Spatial C50 distribution on the \emph{dampenedBase} scene from Hear-Anything-Anywhere dataset, together with its RGB image.}
    \label{fig:haa_heatmap}
\end{figure}

\subsubsection{Sim-to-real Performance on Hear-Anything-Anywhere (HAA)}
To evaluate the sim-to-real ability, we also finetune \tool on HAA as former works. Following the experimental setup in xRIR, four distinct acoustic environments are selected: classroomBase, hallwayBase, complexBase, and dampenedBase. All four environments are partitioned into train and test splits with an 8:2 ratio.
Both the retrained xRIR and \tool are finetuned for 10 epochs for fair comparison. The metrics are shown in Tab~\ref{tab:haa_metrics}, and the way metrics evolve as the reference count $K$ grows is also discussed in Sec.~\ref{sec:k_scaling}.

Our model consistently outperforms all baselines across all four rooms and reference settings (except for Classroom with $4$ reference views), showing that the sim-to-real advantage of \tool is not specific to any single room but reflects a more accurate modeling of real-world clarity and energy decay. Compared to Diff-RIR, our model achieves much better performance without requiring such constraints, highlighting its flexibility and general applicability. 

We pay particular attention to the strongly absorptive \emph{dampenedBase} room, which exhibits substantially longer reverberation times and therefore poses a more challenging scene. In this setting, xRIR's T60 estimates remain excessively large error (around $191\sim254\%$) across all reference counts, indicating poor robustness under challenging configurations. In contrast, our model produces significantly more stable and physically plausible T60 estimates ($46 \sim 49\%$), while also achieving superior C50 and EDT.

To provide a more intuitive view of this challenging scene, Fig.~\ref{fig:haa_heatmap} visualizes the spatial C50 distribution over the \emph{dampenedBase} room. In contrast, EigeNet simultaneously preserves better both global and local consistency, ensuring a more coherent reconstruction of the RIR across the scene.

\subsubsection{Scaling Behavior with Reference Count $K$}
\label{sec:k_scaling}
\begin{figure}[htbp]
    \centering
    \begin{subfigure}{\columnwidth}
        \centering
        \includegraphics[width=\columnwidth]{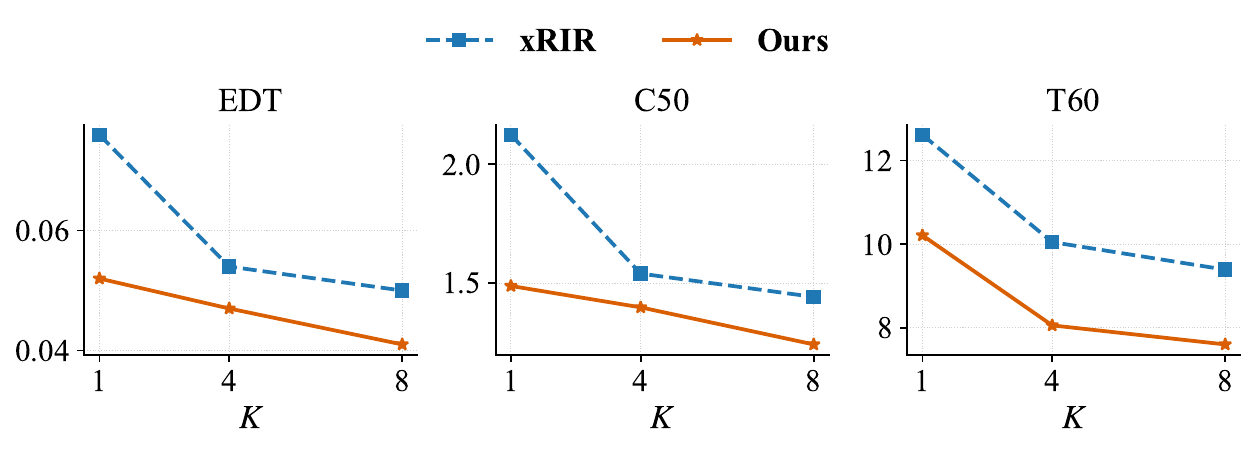}
        \caption{AcousticRooms dataset}
        \label{fig:kscaling_ar}
    \end{subfigure}
    
    \begin{subfigure}{\columnwidth}
        \centering
        \includegraphics[width=\columnwidth]{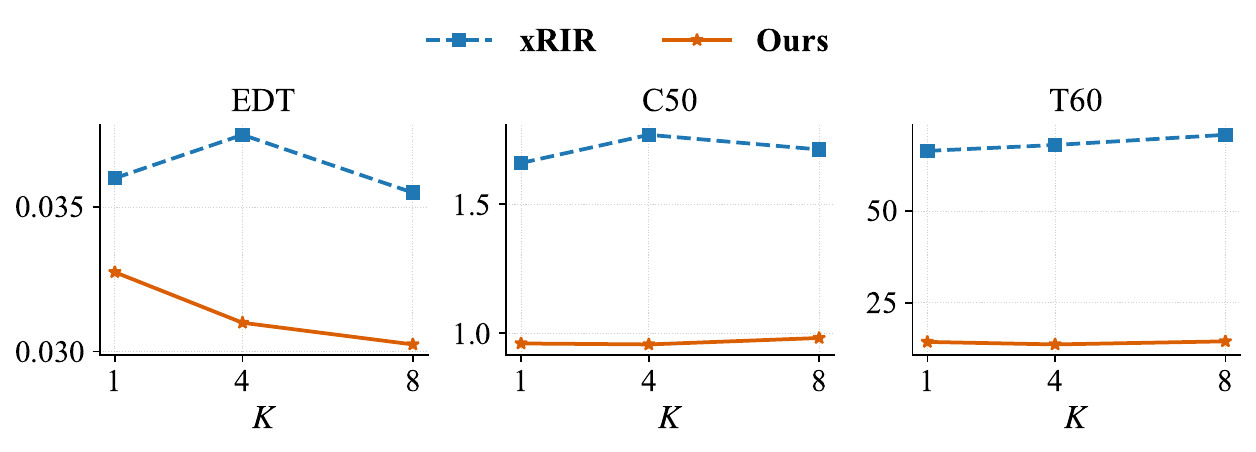}
        \caption{HAA dataset}
        \label{fig:kscaling_haa}
    \end{subfigure}
    
    \caption{Scaling of EDT (s), C50 (dB) and T60 (\%) errors as a function of the number of reference views $K\!\in\!\{1,4,8\}$ (lower is better).}
    \label{fig:kscaling} 
\end{figure}

A central practical question for few-shot \task{} is how the performance scales with the number of reference views $K$. We revisit this question jointly on AR (simulated data) and HAA (real recordings) test splits in Fig.~\ref{fig:kscaling_ar} and Fig.~\ref{fig:kscaling_haa}. For both AR and HAA, we average across rooms to obtain a scene-averaged curve per model.

On AR test split (Fig.~\ref{fig:kscaling_ar}), \tool outperforms xRIR across all reference counts. Further more, \tool already starts from a markedly lower error at $K{=}1$ and the gap over xRIR persists or widens as $K$ decreases. On HAA (Fig.~\ref{fig:kscaling_haa}), the advantages of \tool over xRIR hold still. However, the expected monotonic improvement as $K$ increases weakens. We attribute this trend to potential dataset bias.


\begin{figure}[t]
\centering
\includegraphics[width=\columnwidth]{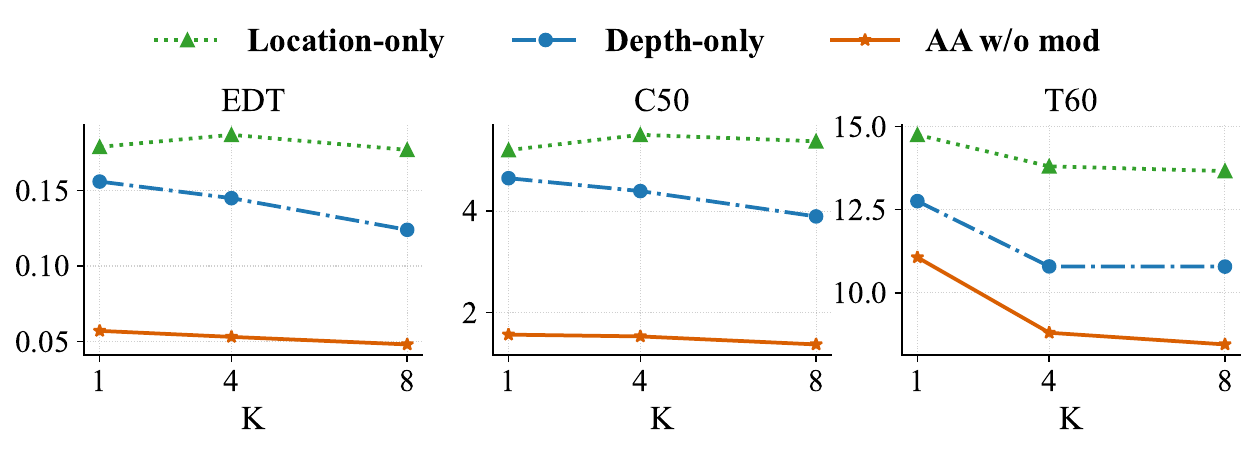}
\caption{Geometric-Inputs ablation on the unseen split of AcousticRooms: broadband EDT (s), C50 (dB) and T60 (\%) errors as a function of reference count $K\!\in\!\{1,4,8\}$ (lower is better).}
\label{fig:ablation_geo_kscale}
\end{figure}

\begin{figure}[t]
\centering
\includegraphics[width=\columnwidth]{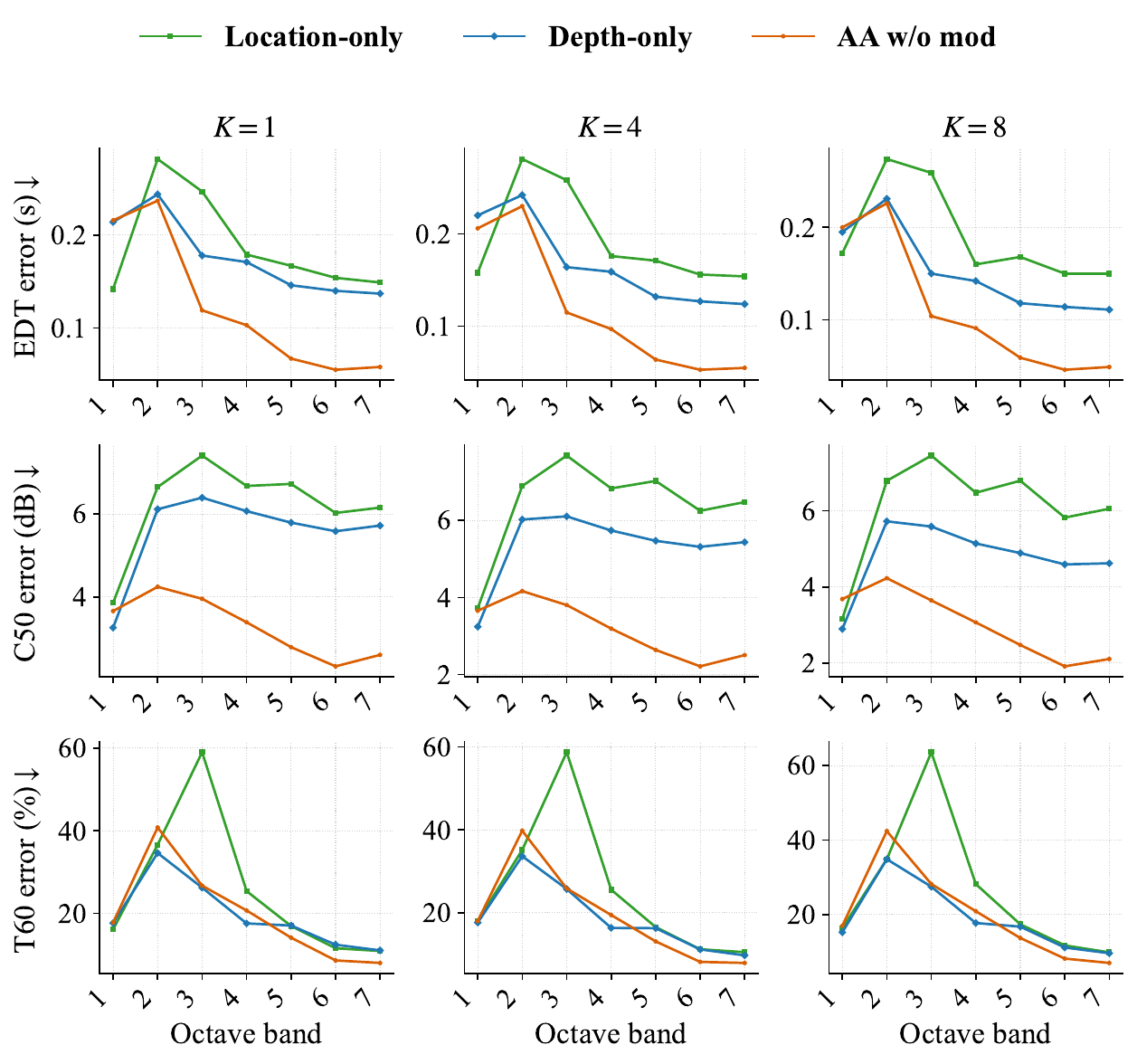}
\caption{Geometric-Inputs ablation on the unseen split of AcousticRooms: multi-octave  EDT (s), C50 (dB) and T60 (\%) errors across octave bands at $K\!\in\!\{1,4,8\}$ (lower is better).}
\label{fig:ablation_geo_oct}
\end{figure}

\subsection{Ablation Study}
To evaluate the individual contributions of each component within our proposed framework, we conduct three sets of ablation experiments.  
Specifically, the first set investigates the impact of utilizing different combinations of geometric inputs on the reconstruction performance. 
The second set explores the advantages of the proposed Alternate-attention mechanism over conventional Self-attention and Cross-attention architectures.
The third set verifies the necessity of the proposed geometry-informed modulation block.

To provide a comprehensive analysis of the ablation configurations, we further conduct a multi-octave band evaluation by calculating the corresponding EDT, C50, and T60 errors on the test split of the AR dataset. The multi-octave bands span from the band centered at 63 Hz up to 4~kHz.

\begin{figure}[t]
\centering
\includegraphics[width=\columnwidth]{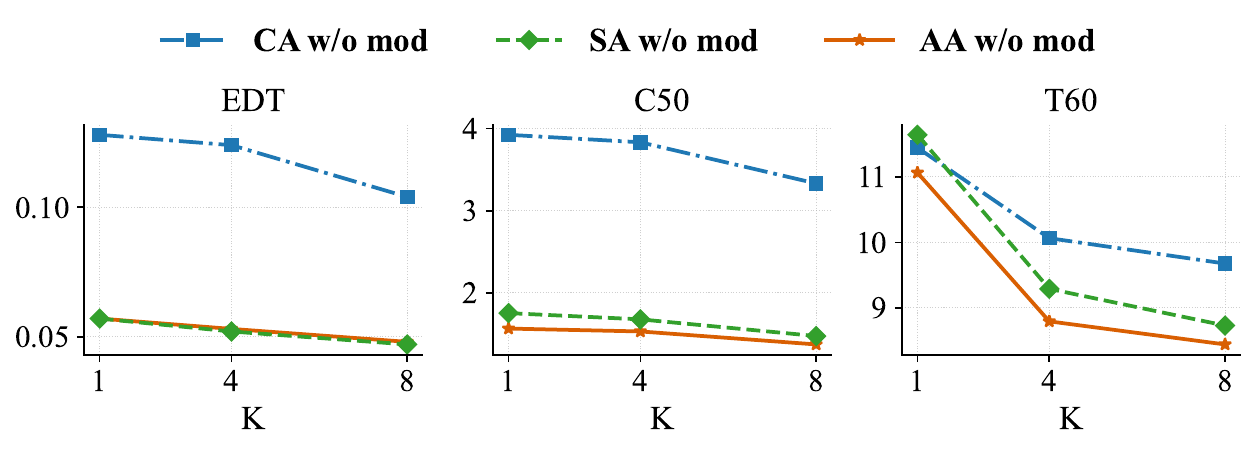}
\caption{Attention-Mechanism ablation on the unseen split of AcousticRooms: broadband EDT (s), C50 (dB) and T60 (\%) errors as a function of reference count $K\!\in\!\{1,4,8\}$ (lower is better).}
\label{fig:ablation_attn_kscale}
\end{figure}

\begin{figure}[t]
\centering
\includegraphics[width=\columnwidth]{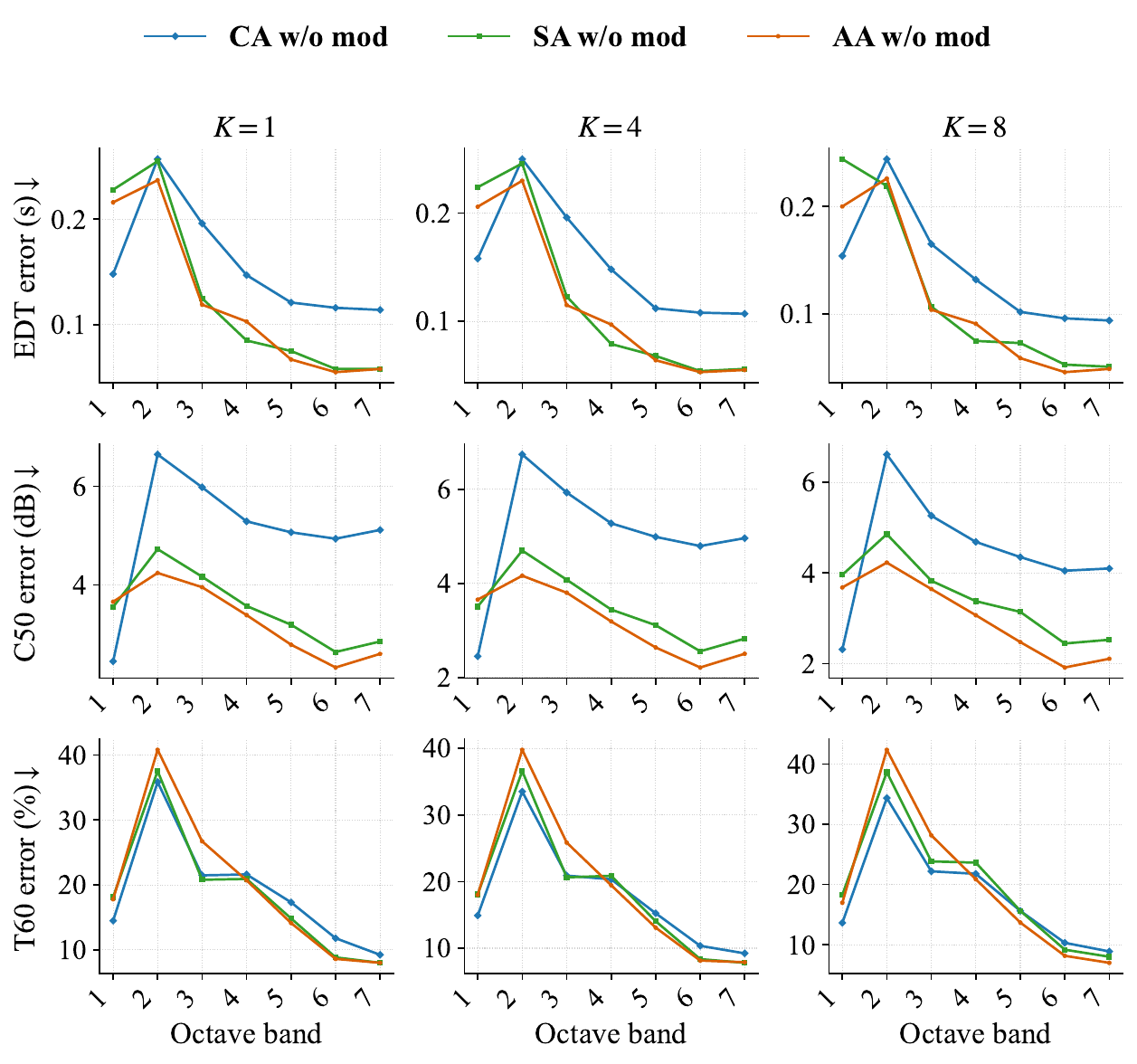}
\caption{Attention-Mechanism ablation on the unseen split of AcousticRooms: multi-octave EDT, C50, and T60 errors across octave bands at $K\!\in\!\{1,4,8\}$.}
\label{fig:ablation_attn_oct}
\end{figure}

\subsubsection{Influence of the Geometric Inputs}
Firstly, we investigate the necessity of providing comprehensive geometric context by training models with isolated geometric input, specifically using either only panoramic depth maps (\emph{Depth-only}) or only coordinate locations (\emph{Location-only}). To ensure a fair comparison and strictly isolate the impact of the geometric inputs, both of these variants are trained while bypassing the modulation block and directly using the 
$T_0$ as the acoustic tokens for the target view in stead. Therefore, their performances are directly compared against the \emph{AA w/o mod} baseline.

As observed in Fig~\ref{fig:ablation_geo_kscale}, omitting either geometric modality leads to severe performance degradation on broadband metrics across all reference counts. These degradations are particularly pronounced in the EDT and C50, which is consistent with the findings reported in FLAC~\cite{FLAC}.
The multi-octave band analysis in Fig.~\ref{fig:ablation_geo_oct} further validate the necessity of the compressive geometric inputs. 

\begin{table*}[htbp]
\centering
\caption{Masking probe on the unseen split of AcousticRooms.}
\label{tab:probe}

\begin{subtable}{\textwidth}
    \centering
    \caption{Masking Geometric Tokens: we feed all 8 reference acoustic tokens but keep only the first $n_{\text{geo}}\!\in\!\{0,1,4\}$ reference views' geometric tokens visible, attention-masking the rest. The acoustic tokens of all reference views and all tokens of target view (geometric and acoustic tokens) are visible.}
    \label{tab:mask_geo_token}
    \begin{tabular}{cccccccccc}
    \toprule
    & \multicolumn{3}{c|}{CA w/o mod} & \multicolumn{3}{c|}{SA w/o mod} & \multicolumn{3}{c}{AA w/o mod} \\
    \midrule
    $n_{\text{geo}}$ & EDT(s) $\downarrow$ & C50(dB) $\downarrow$ & T60(\%) $\downarrow$ & EDT(s) $\downarrow$ & C50(dB) $\downarrow$ & T60(\%) $\downarrow$ & EDT(s) $\downarrow$ & C50(dB) $\downarrow$ & T60(\%) $\downarrow$ \\
    \midrule
    0 &0.11&3.444&9.908& 0.051 & 1.710 & 8.429 & 0.153 & 4.712 & 11.577 \\
    1 &0.106&3.381&9.855& 0.050 & 1.667 & 8.370 & 0.059 & 1.730 & 8.422 \\
    4 &0.105&3.354&9.769& 0.048 & 1.574 & 8.526 & 0.053 & 1.547 & 8.098 \\
    \bottomrule
    \end{tabular}
\end{subtable}

\vspace{0.4cm} 

\begin{subtable}{\textwidth}
    \centering
    \caption{Masking Acoustic Tokens: we feed all 8 reference acoustic tokens but keep only the first $n_{\text{ac}}\!\in\!\{0,1,4\}$ reference views' geometric tokens visible, attention-masking the rest. The geometric tokens of all reference views and all tokens of target view (geometric and acoustic tokens) are visible.}
        \label{tab:mask_ac_token}
    \begin{tabular}{cccccccccc}
    \toprule
        & \multicolumn{3}{c|}{CA w/o mod} & \multicolumn{3}{c|}{SA w/o mod} & \multicolumn{3}{c}{AA w/o mod} \\
    \midrule
    $n_{\text{ac}}$ & EDT(s) $\downarrow$ & C50(dB) $\downarrow$ & T60(\%) $\downarrow$ & EDT(s) $\downarrow$ & C50(dB) $\downarrow$ & T60(\%) $\downarrow$ & EDT(s) $\downarrow$ & C50(dB) $\downarrow$ & T60(\%) $\downarrow$ \\
    \midrule
    0&1.398&13.416&25.023&1.342&12.768&29.440&1.318&11.533&28.817\\
    1&0.115&3.465&10.898&0.119&2.676&16.785&0.121&2.684&17.012\\
    4&0.106&3.35&9.671&0.058&1.637&9.850&0.060&1.562&9.989\\
    \bottomrule
    \end{tabular}
\end{subtable}

\vspace{0.4cm} 

\begin{subtable}{\textwidth}
\centering
\caption{Complere Context: all geometric and acoustic tokens are visible for all views. Results are reported under varying reference counts $K\!\in\!\{1,4,8\}$.}
\label{tab:ablation_attn}
\begin{tabular}{cccccccccc}
\toprule
 & \multicolumn{3}{c|}{CA w/o mod} & \multicolumn{3}{c|}{SA w/o mod} & \multicolumn{3}{c}{AA w/o mod} \\
\midrule
$K$ & EDT(s) $\downarrow$ & C50(dB) $\downarrow$ & T60(\%) $\downarrow$ & EDT(s) $\downarrow$ & C50(dB) $\downarrow$ & T60(\%) $\downarrow$ & EDT(s) $\downarrow$ & C50(dB) $\downarrow$ & T60(\%) $\downarrow$ \\
\midrule
1 &0.128&3.924&11.444& 0.057 & 1.755 & 11.641& 0.057 & 1.567 & 11.063 \\
4 &0.124&3.833&10.064& 0.052 & 1.679 & 9.291& 0.053 & 1.532 & 8.795 \\
8 &0.104&3.332&9.677& 0.047 & 1.476 & 8.731& 0.048 & 1.374 & 8.443  \\
\bottomrule
\end{tabular}
\end{subtable}

\end{table*}
    
\subsubsection{Impact of Different Attention Mechanisms}
\label{sec:attention-ablation}

We compare \emph{Alternate-attention} (\emph{AA w/o mod}) against two standard attention mechanisms: 
(i) \emph{Cross-attention (CA w/o mod) }, where references serve as keys and values for the target view, and (ii) \emph{Self-attention (SA w/o mod )}, where all tokens are flattened into a single sequence and all the tokens can attend to each other. All the variants are trained without the modulation block.

To ensure a fair and rigorous comparison, we keep all training configuration the same with the sole exception of the attention mechanism. 
Specifically, both the Self-attention and Cross-attention configurations are implemented with 12 Transformer layers and the same number of attention heads as our 6 Alternate-attention block backbone (which also amounts to 12 attention layers in total).

\begin{figure}[htbp]
    \centering
    
    \begin{subfigure}[b]{\columnwidth}
        \centering
        \includegraphics[width=\textwidth]{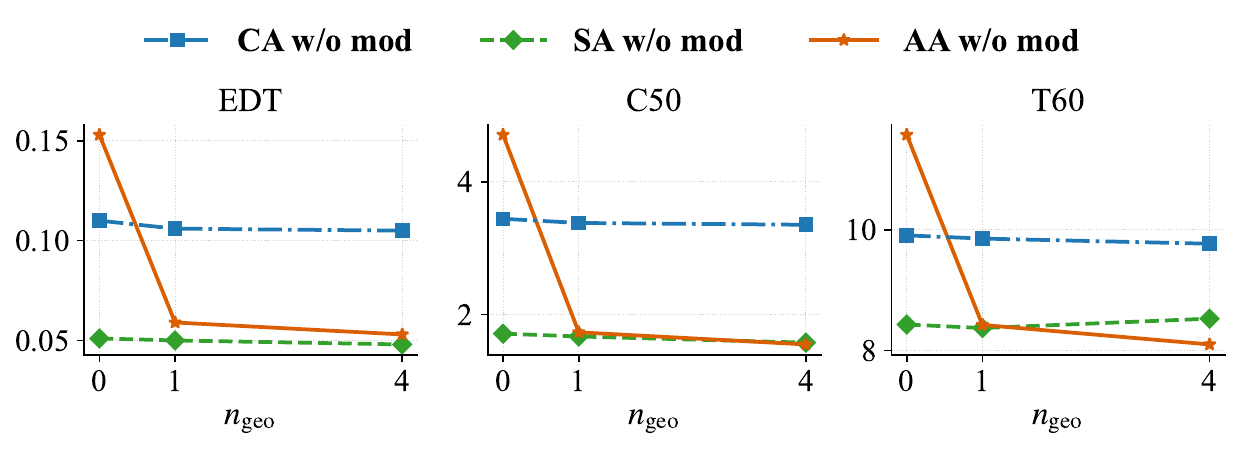}
        \caption{Geometric-token masking probe on AcousticRooms. All 8 reference acoustic tokens are always kept; only the first $n_{\text{geo}}$ reference geometric tokens are kept, the rest are attention-masked.}
        \label{fig:probe_geo_masking}
    \end{subfigure}
    
    \vspace{5pt}
    
    \begin{subfigure}[b]{\columnwidth}
        \centering
        \includegraphics[width=\textwidth]{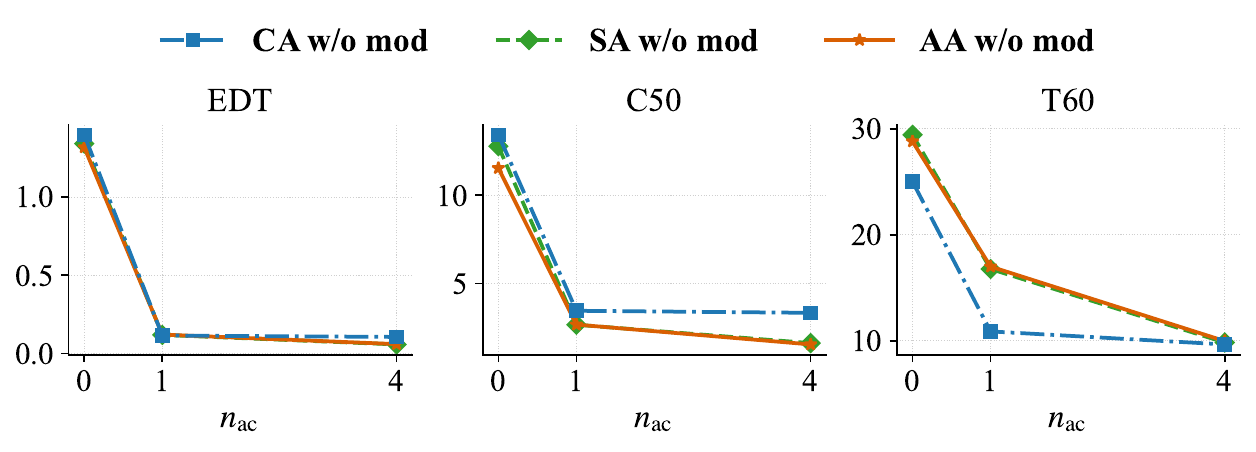}
        \caption{Acoustic-token masking probe on AcousticRooms. All 8 reference geometric tokens are always kept; only the first $n_{\text{ac}}$ reference acoustic tokens are kept, the rest are attention-masked.}
        \label{fig:probe_ac_masking}
    \end{subfigure}
    
    \caption{Acoustic and geometric token masking probes on AcousticRooms.}
    \label{fig:probe}
\end{figure}

As shown in Fig~\ref{fig:ablation_attn_kscale}, Cross-attention yields the most significant performance degradation across all metrics.
While both Self-attention and Alternate-attention maintain a highly comparable EDT, Alternate-attention demonstrates clear better performance on C50 and T60. 

Fig~\ref{fig:ablation_attn_oct} shows that Alternate-attention exhibits minor advantages to Self-attention and far exceeds Cross-attention on EDT; consistently outperforms on all octave bands across C50; slight advantages within the mid-to-high octave bands on T60. 

Because the broadband performance gap between Alternate-attention and Self-attention is moderate, we conduct a deeper experiment to explore the differenct multi-view multi-modal reasoning patterns among Cross-attention, Self-attention and Alternate-attention while predicting the target RIR.
The experiment includes two inference-time probes on the Cross-attention, Self-attention and Alternate-attention. 

For the first one, we feed eight reference views and keep only the first $n_{\text{geo}}\!\in\!\{0,1,4\}$ reference views' geometric tokens visible, while the remaining views' geometric tokens are masked in the attention calculation. The acoustic tokens of all reference views and all tokens of target view (geometric and acoustic tokens) are always visible.
The second setting takes the opposite approach---masking the first $n_{\text{ac}}\!\in\!\{0,1,4\}$ reference views' acoustic tokens while keeping the geometric tokens visible.
By doing this, we can directly evaluate how the model reacts to the absence of certain modality.
We also show the metrics of normal inference (all geometric and acoustic tokens are visible for all views) with $K$ references as a comparison.
The trends of the evaluation metrics are illustrated as a function of $n_{geo}$ and $n_{ac}$ in Fig.~\ref{fig:probe} and the specific metrics are reported in Tab.~\ref{tab:probe}. From the two probes, we discover the different nature between Cross-attention,Self-attention and Alternate-attention.

\emph{1. Cross-attention and Self-attention's reasoning ability depends on interpolating over reference acoustic tokens.} We conclude that from three observations:
\begin{itemize}
    \item The improvement is marginal with the growth of $n_{geo}$ but remarkable with the growth of $n_{ac}$;
    \item The metrics on $n_{\text{geo}}{=}0$ are not collapsed compared to that on $K=8$;
    \item Comparing $n_{\text{geo}}{=}1/4$ with $K=1/4$, we find that the metrics are even better.
\end{itemize}
The first two observations collectively suggest that Cross-attention and Self-attention exhibits a minimal dependence on geometric tokens, while being heavily reliant on acoustic tokens. Our third observation further suggests that: when $n_{\text{geo}}{=}K$, because both have access to an identical complete set of geometric and acoustic tokens from reference views, the extra acoustic tokens cause the final improvement on the metrics. These observations together confirms either Cross-attention or Self-attention is fundamentally a soft interpolator over acoustic tokens.

\emph{2. Alternate-attention's reasoning ability depends on multi-view multi-modal context.} We conclude that from three observations:
\begin{itemize}
    \item metrics maintain continuous improvement with more complete context of reference views;
    \item the metrics are all collapsed when $n_{\text{geo}}{=}0$ and $n_{\text{ac}}{=}0$;
    \item the metrics on $n_{\text{geo}}{=}1/4$ are worse than on $K=1/4$.
\end{itemize}
The first two observations preliminarily establish that Alternate-attention cannot work properly by relying solely on geometric or acoustic tokens; instead, it necessitates a comprehensive multi-view multi-modal context to achieve optimal performance. This perspective is further substantiated by the third observation, which confirms that Alternate-attention does not operate as a mere soft interpolator, because the extra acoustic tokens do no help to the overall metrics.

It's worth mentioning that T60 alone breaks this pattern and Cross-attention, Self-attention and Alternate-attention depend mostly on acoustic tokens when regressing T60.
We propose a plausible explanation for this phenomenon.
Due to the physical structure of the RIR --- $T_{60}$ is a diffuse-field quantity and can be estimated roughly by \emph{Sabine Equation}~\cite{roomacoustics}:
\begin{equation}
\label{eq:sabin equation}
T_{60}\!\propto\!V/(c\,\bar{\alpha}\,S)    
\end{equation}
where $V$ is the room volume, $c$ is the speed of sound, $\bar{\alpha}$ is the average absorption coefficient, and $S$ is the total surface area. 
T60 is weakly position-dependent, so the model can just leverage the existing reference acoustic tokens to infer the T60 of the novel view RIR. 

\begin{table}[t]
    \centering
    \caption{Ablation study on different target choices for the modulation block on the unseen split of AcousticRooms under $K \in \{1,4,8\}$. We compare our proposed 7-band multi-octave power spectrum mapping against variants without power spectrum loss ( ``Var.1'') and the full STFT power spectrum (``Var.2''). The best result in each $K$ block is \textbf{bold}. The metrics of ``AA w/o mod'' is also listed as a comparison.}
    \label{tab:ablation_spectrum_loss}
    \resizebox{0.9\columnwidth}{!}{%
    \begin{tabular}{lcccc}
    \toprule
    \textbf{Method} & \textbf{$K$} & \textbf{EDT (s)} $\downarrow$ & \textbf{C50 (dB)} $\downarrow$ & \textbf{T60} $\downarrow$ \\
    \midrule
    AA w/o mod & 1 & 0.057 & 1.567 & 11.063 \\
    Var.1 & 1 & 0.053 & 1.605 & 10.958 \\
    Var.2 & 1 & 0.054 & 1.515 & 12.079 \\
    \textbf{Ours}  & 1 & \textbf{0.052} & \textbf{1.488} & \textbf{10.213} \\
    \midrule
    AA w/o mod & 4 & 0.053 & 1.532 & 8.795 \\
    Var.1 & 4 & 0.049 & 1.540 & 8.759 \\
    Var.2 & 4 & 0.049 & 1.419 & 9.777 \\
    \textbf{Ours}  & 4 & \textbf{0.047} & \textbf{1.398} & \textbf{8.061} \\
    \midrule
    AA w/o mod & 8 & 0.048 & 1.374 & 8.443 \\
    Var.1 & 8 & 0.044 & 1.357 & 8.512 \\
    Var.2 & 8 & 0.043 & 1.260 & 9.902 \\
    \textbf{Ours}  & 8 & \textbf{0.041} & \textbf{1.242} & \textbf{7.605} \\
    \bottomrule
    \end{tabular}%
    }
\end{table}

\begin{table}[t]
    \centering
    \caption{Generalization of the modulation block across attention mechanisms on the unseen split of AcousticRooms. For each attention type Alternate-Attention/Self-attention/Cross-attention (AA/SA/CA), we report ``w/ mod'' and ``w/o mod'' variants under $K\!\in\!\{1,4,8\}$ (Ours model is AA w/ mod). The best result in each $K$ block is \textbf{bold}.}
    \label{tab:ablation_mod}
    \resizebox{0.95\columnwidth}{!}{%
    \begin{tabular}{lcccc}
    \toprule
    \textbf{Method} & \textbf{$K$} & \textbf{EDT (s)} $\downarrow$ & \textbf{C50 (dB)} $\downarrow$ & \textbf{T60} $\downarrow$ \\
    \midrule
    CA w/o mod                     & 1 & 0.128 & 3.924 & 11.444 \\
    CA w/ mod          & 1 & 0.060 & 1.595 & 11.154 \\
    SA w/o mod                     & 1 & 0.057 & 1.755 & 11.641 \\
    SA w/ mod          & 1 & 0.053 & 1.567 & 12.037 \\
    AA w/o mod            & 1 & 0.057 & 1.567 & 11.063 \\
    \textbf{AA w/ mod}  & 1 & \textbf{0.052} & \textbf{1.488} & \textbf{10.213} \\
    \midrule
    CA w/o mod                     & 4 & 0.124 & 3.833 & 10.064 \\
    CA w/ mod          & 4 & 0.049 & 1.424 & 8.904 \\
    SA w/o mod                     & 4 & 0.052 & 1.679 & 9.291 \\
    SA w/ mod          & 4 & 0.048 & 1.498 & 9.388 \\
    AA w/o mod           & 4 & 0.053 & 1.532 & 8.795 \\
    \textbf{AA w/ mod}  & 4 & \textbf{0.047} & \textbf{1.398} & \textbf{8.061} \\
    \midrule
    CA w/o mod                     & 8 & 0.104 & 3.332 & 9.677 \\
    CA w/ mod          & 8 & 0.044 & 1.302 & 8.513 \\
    SA w/o mod                     & 8 & 0.047 & 1.476 & 8.731 \\
    SA w/ mod          & 8 & 0.042 & 1.283 & 8.902 \\
    AA w/o mod             & 8 & 0.048 & 1.374 & 8.443 \\
    \textbf{AA w/ mod}  & 8 & \textbf{0.041} & \textbf{1.242} & \textbf{7.605} \\
    \bottomrule
    \end{tabular}%
    }
\end{table}

\begin{figure}[t]
\centering
\includegraphics[width=\columnwidth]{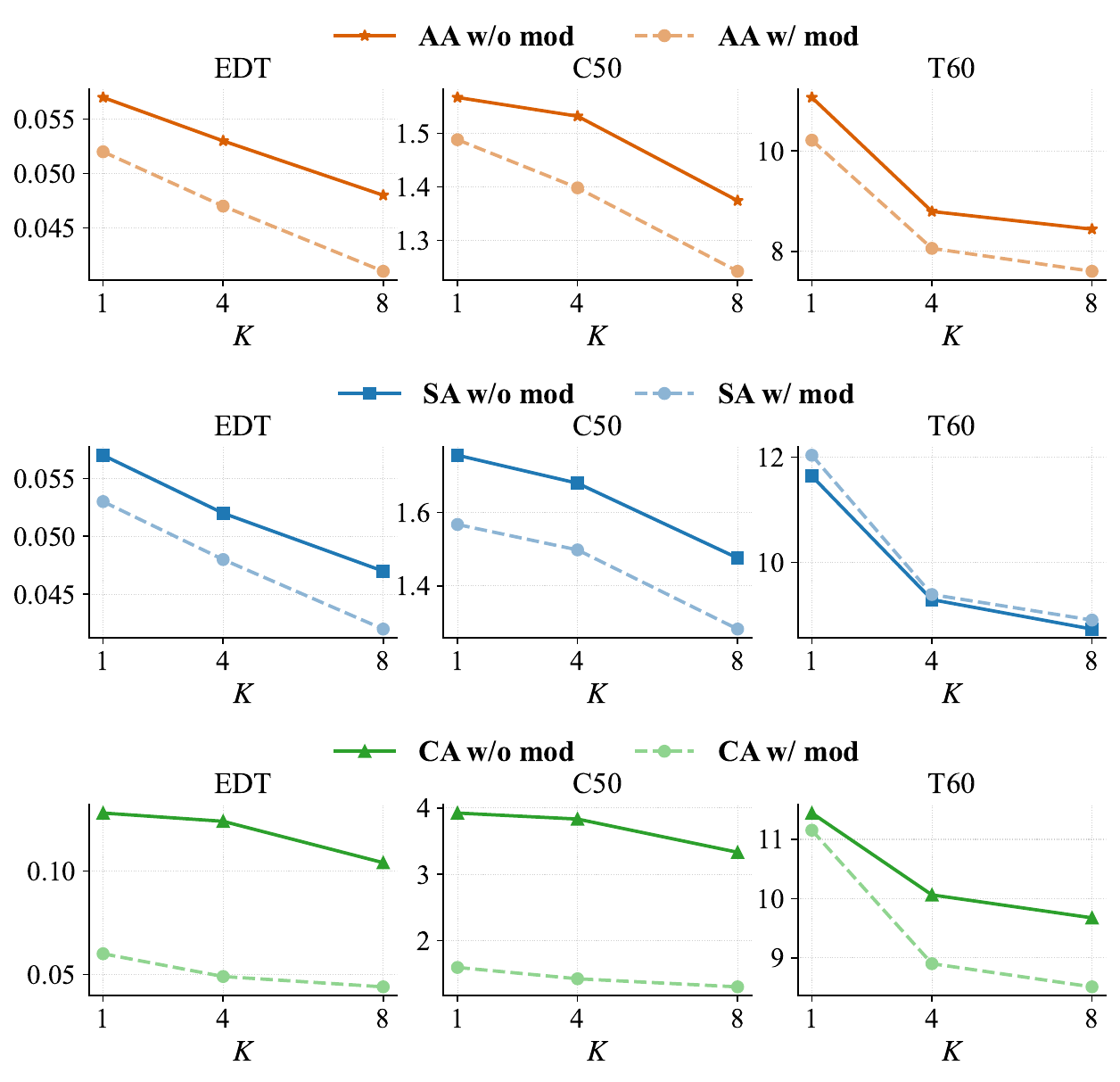}
\caption{Per-metric comparison of all six variants across $K\!\in\!\{1,4,8\}$. Rows group attention types (AA/SA/CA); columns show EDT, C50, and T60 errors.}
\label{fig:ablation_mod}
\end{figure}

\subsubsection{Impact of the Modulation Block}

Firstly, we investigate the influence of different power spectrum targets on RIR prediction. Considering our proposed target being the 7-band multi-octave power spectrum, this ablation study evaluates two alternative variants: (1) omitting the power spectrum loss (denoted by ``Var.1''), and (2) a variant employing the Short-Time Fourier Transform power spectrum (denoted by ``Var.2'') computed with a 1024-sample window, yielding 513 frequency bins.

The results are illustrated in Tab.~\ref{tab:ablation_spectrum_loss}. 
First, the similar performance between Var.1 and AA w/o mod demonstrates that mere modulation block without the power spectrum loss as the explicit task-specific supervision fails to yield meaningful geometric-acoustic representation gains.
Second, employing the full STFT spectrum (Var.2) improves EDT and C50 but weaken the T60, as the dense 513 frequency bins contains excessive acoustic details being a hard target to regress, whereas our 7-band multi-octave reduction provides a cleaner, physics-informed signal.

Furthermore, to validate the fundamental and generalizable utility of the modulation block in RIR prediction, we evaluate it across three attention designs: our proposed Alternate-attention (AA), standard Self-attention (SA), and Cross-attention (CA). For each design, we train two variants---with and without the modulation block---under identical training configurations.

The quantitative results in Tab.~\ref{tab:ablation_mod} and the visual comparison in Fig.~\ref{fig:ablation_mod} together reveal two key findings.

\textbf{1. The modulation block provides a consistent and substantial gain across all attention mechanisms.} Within every attention pair (AA, SA, CA) and across all $K$, the variant with modulation block uniformly outperforms its counterpart. The improvement is most dramatic for CA, bringing the previously worst-performing attention design to a competitive level. For SA, the modulation block also yields clear gains, confirming that the benefit is not exclusive to AA.

\textbf{2. Ours (Alternate-attention + modulation block) achieves the best overall performance.} While the modulation block universally improves all attention types, the combination of Alternate-attention and modulation block consistently achieves the lowest errors across all metrics and reference counts.

\section{Conclusions}
In this paper, we proposed \tool for few-shot \task{} that integrates data-driven modeling with geometry-informed physical priors. 
To capture the complex spatial-temporal dependencies of RIRs, we introduced a Cross-view Alternate-attention Transformer. Furthermore, inspired by acoustic ray tracing, we designed a geometry-informed modulation block under a multi-task learning paradigm.
Extensive experiments demonstrate that \tool achieves state-of-the-art performance across both simulated and real-world datasets.
Beyond benchmark performance, we design a series of experiments to provide empirical evidence for the interpretability and necessity of our proposed design, aiming to uncover the underlying mechanisms that drive the observed performance gains.


\bibliographystyle{IEEEtran}
\bibliography{mybib}

\end{document}